\documentclass[nojss,noheadings]{jss}
\usepackage{amsmath,amstext,booktabs,dcolumn,thumbpdf}

\newcommand{\fct}[1]{\code{#1()}}

\DefineVerbatimEnvironment{Sinput}{Verbatim}{}
\setkeys{Gin}{width=\textwidth}

\title{Visualizing Count Data Regressions Using Rootograms}

\author{Christian Kleiber\\Universit\"at Basel
  \And Achim Zeileis\\Universit\"at Innsbruck}
\Plainauthor{Christian Kleiber, Achim Zeileis}

\Abstract{
  The rootogram is a graphical tool associated with the work of J.
  W. Tukey that was originally used for assessing goodness of fit of univariate
  distributions. Here we extend the rootogram to regression models and show that
  this is particularly useful for diagnosing and treating issues such as overdispersion
  and/or excess zeros in count data models.
  We also introduce a weighted version of the rootogram that can be applied out of sample
  or to (weighted) subsets of the data, e.g., in finite mixture models. 
  An empirical illustration revisiting a well-known data set from ethology is included, for which a  
  negative binomial hurdle model is employed. 
  Supplementary materials providing two further illustrations are available online: the first, using data from public health, employs a two-component
  finite mixture of negative binomial models, the second, using data from finance, involves underdispersion.
  An \proglang{R} implementation of our tools is available in the \proglang{R}~package \pkg{countreg}. 
  It also contains the data and replication code.
}


\Address{
  Christian Kleiber\\
  Faculty of Business and Economics\\
  
  Universit\"at Basel\\  
  Peter Merian-Weg 6\\
  4002 Basel, Switzerland\\
  E-mail: \email{Christian.Kleiber@unibas.ch}\\
  URL: \url{http://wwz.unibas.ch/kleiber/}\\

  Achim Zeileis\\
  Department of Statistics\\
  Faculty of Economics and Statistics\\
  Universit\"at Innsbruck\\
  Universit\"atsstr.~15\\
  6020 Innsbruck, Austria\\
  E-mail: \email{Achim.Zeileis@R-project.org}\\
  URL: \url{http://eeecon.uibk.ac.at/~zeileis/}
}

\usepackage{fancyhdr}
\fancypagestyle{plain}{
\fancyhf{}
\fancyfoot[LO,LE]{\small This is a preprint of an article accepted for publication in \textit{The American Statistician}.
  \doi{10.1080/00031305.2016.1173590} \\
  Copyright~{\copyright} 2016 American Statistical Association. \url{http://www.tandfonline.com/r/tas}.
  }
}

\begin{document}

\section{Introduction} \label{sec:introduction}

The area of count data regression has experienced rapid growth over the last two decades. More often than 
not, the standard Poisson model from the generalized linear model (GLM) toolbox does not suffice in empirical work. Specifically, many 
data sets are plagued by some form of overdispersion, often resulting from unobserved 
heterogeneity that can potentially be handled by, e.g., models with additional shape parameters such as the negative 
binomial distribution or from an excess of zeros for which hurdle and zero-inflation models are available
\citep{rootograms:Mullahy:1986, rootograms:Lambert:1992}. While various diagnostic tests of dispersion are also available 
-- see, e.g., \cite{rootograms:Cameron+Trivedi:1990} or \cite{rootograms:Dean:1992} for some popular tests 
and \cite{rootograms:Cameron+Trivedi:2013} for an overview -- they typically only identify general issues with 
model fit and rarely provide clear indications regarding the source of the problems. Suitable graphical tools 
can guide the search for more appropriate specifications, thereby supplementing and enhancing more formal approaches. 

If count data regressions are visualized at all, this is 
currently mainly done in the form of barplots of observed and expected frequencies; see, e.g., Figures~3.1 and~6.4 in \cite{rootograms:Cameron+Trivedi:2013} for examples and also 
Figure~\ref{fig:CrabSatellites-comparison} below. In the present paper, we explore the use of
rootograms for assessing the fit. Rootograms  are associated with the work of John W. Tukey on exploratory data analysis (EDA) and 
statistical graphics, culminating in \cite{rootograms:Tukey:1977}. However, rootograms do not figure 
prominently there. Instead, early applications, all confined to continuous data, appear in selected 
contributions to collected volumes and conference proceedings \citep{rootograms:Tukey:1965,
rootograms:Tukey:1972}, which were often not easily available prior to the publication of Tukey's 
collected works in the 1980s. Nonetheless, the ideas pertaining to rootograms were known in some circles 
at an early stage \citep{rootograms:Healy:1968}, and an early paper popularizing the concept is
\cite{rootograms:Wainer:1974}. For further information on the history of statistical graphics we refer to
\cite{rootograms:Friendly+Denis:2001}.

The following section introduces a generalized version of the rootogram for regression models
(as opposed to univariate distributions) and allowing for weights that can be applied to
new data or (weighted) subsamples of a data set. This is useful for assessing in-sample fits as well as
out-of-sample predictions and also for situations with survey weights or model-based weights. 
Several styles of the rootogram, namely standing, hanging, and suspended versions, are 
briefly described. We also provide some guidelines for interpretation using simulated data. 
Section~\ref{sec:example} provides an empirical example, presenting a case
where a hurdle model adjusts for excess zeros and also for overdispersion, while the final 
section~\ref{sec:disc} discusses how rootograms could be included in routine applications of count data regressions. 
In supplementary materials, we present two further examples, one involving a finite mixture model requiring 
the rootogram version with model-based weights mentioned above, the other involving underdispersed data.

All analyses are run in \proglang{R} \citep{rootograms:R:2016}, and we briefly describe an 
implementation of our tools in the \proglang{R} package \pkg{countreg} in an appendix.

\section{Rootograms} \label{sec:rootograms}

Given observations $y_i$ ($i = 1, \dots, n$) we want to assess the goodness of
fit of some parametric model $F(\cdot; \alpha_i)$, with corresponding density or
probability mass  function $f(\cdot; \alpha_i)$. 
For classic rootograms \citep[see e.g.,][Chapter~2]{rootograms:Friendly:2000}
the parameter vector $\alpha_i$ is the same for all observations $i = 1, \dots, n$.
Here, we allow it to be observation-specific, e.g., through dependence on some covariates
$x_i$ -- a leading case being the GLM with $\alpha_i = g(x_i^\top \beta)$
for some monotonic function $g(\cdot)$. In practice, these parameters are
typically unknown and have to be estimated from data. Hence, in the following we
assume that we have fitted parameters $\hat \alpha_i$ where estimation may have
been carried out on the same observations $i = 1, \dots, n$ (i.e., corresponding
to an in-sample assessment) or on a different data set (i.e., out-of-sample
evaluation). The estimation procedure itself may be fully parametric or
semiparametric etc.\ as long as it yields fitted parameters $\hat \alpha_i$ for
all observations of interest. 

To judge the goodness of fit of a model with estimated parameters $\hat
\alpha_i$ to observations $y_i$ ($i = 1, \dots, n$), a natural idea is to assess
whether observed frequencies match expected frequencies from the model. In the case
of discrete observations frequencies for the observations themselves could be
considered while somewhat more generally frequencies for intervals of
observations may be used. 
Tukey's original work often considered goodness of fit to the normal
distribution on the basis of binned observations, see, e.g., his example
involving the heights of 218 volcanos \citep{rootograms:Tukey:1972}. In this paper, we
focus on discrete distributions. 

For assessing the goodness of fit in regression models, practitioners routinely
check some type of residuals, i.e., (weighted) deviations of the observations
$y_i$ from the corresponding predicted means. However, this focuses on the first
moment of the fitted distribution only while for count data, which are
non-negative and typically skewed, further aspects of the distribution are also
of interest. Relevant aspects include the amount of (over-)dispersion, skewness
(or further aspects of shape), and whether there are excess zeros. Hence,
it is natural to consider observed and expected values for a range of counts
$0, 1, 2, \dots$ in order to assess the entire fitted distribution. 

Specifically, in the case of count data with
possible outcomes $j = 0, 1, 2, \dots$, the observed and expected frequencies
for each integer $j$ are given by
\begin{eqnarray*}
  \text{obs}_j & = & \sum_{i = 1}^n I(y_i = j) , \\
  \text{exp}_j & = & \sum_{i = 1}^n f(j; \hat \alpha_i) ,
\end{eqnarray*}
where $I(\cdot)$ is an indicator variable. More generally, one can use a set of breaks $b_0, b_1, b_2, \dots$ that span
(a suitable subset of) the support of $y$. Here, we additionally also allow for observation-specific weights $w_i$
 ($i = 1, \dots, n$), the observed and expected frequencies are then given by 
\begin{eqnarray*}
  \text{obs}_j & = & \sum_{i = 1}^n w_i \, I(y_i \in (b_{j}, b_{j + 1}]) , \\
  \text{exp}_j & = & \sum_{i = 1}^n w_i \, \{ F(b_{j + 1}; \hat \alpha_i) - F(b_{j}; \hat \alpha_i) \} .
\end{eqnarray*}
The weights are needed for survey data and also for situations with model-based weights.  
For example, the latter may represent class membership in mixture models, a case that is
relevant in one of our supplementary examples.

\subsection{Styles of Rootograms} \label{subsec:styles}

The rootogram compares observed and expected values graphically by plotting histogram-like
rectangles or bars for the observed frequencies and a curve for the fitted frequencies, 
all on a square-root scale. The square roots rather than the untransformed observations are
employed to approximately adjust for scale differences across the $j$ values or intervals.
Otherwise, deviations would only be visible for $j$'s with large observed/expected
frequencies. 

\begin{figure}[t!]
\centering
\includegraphics{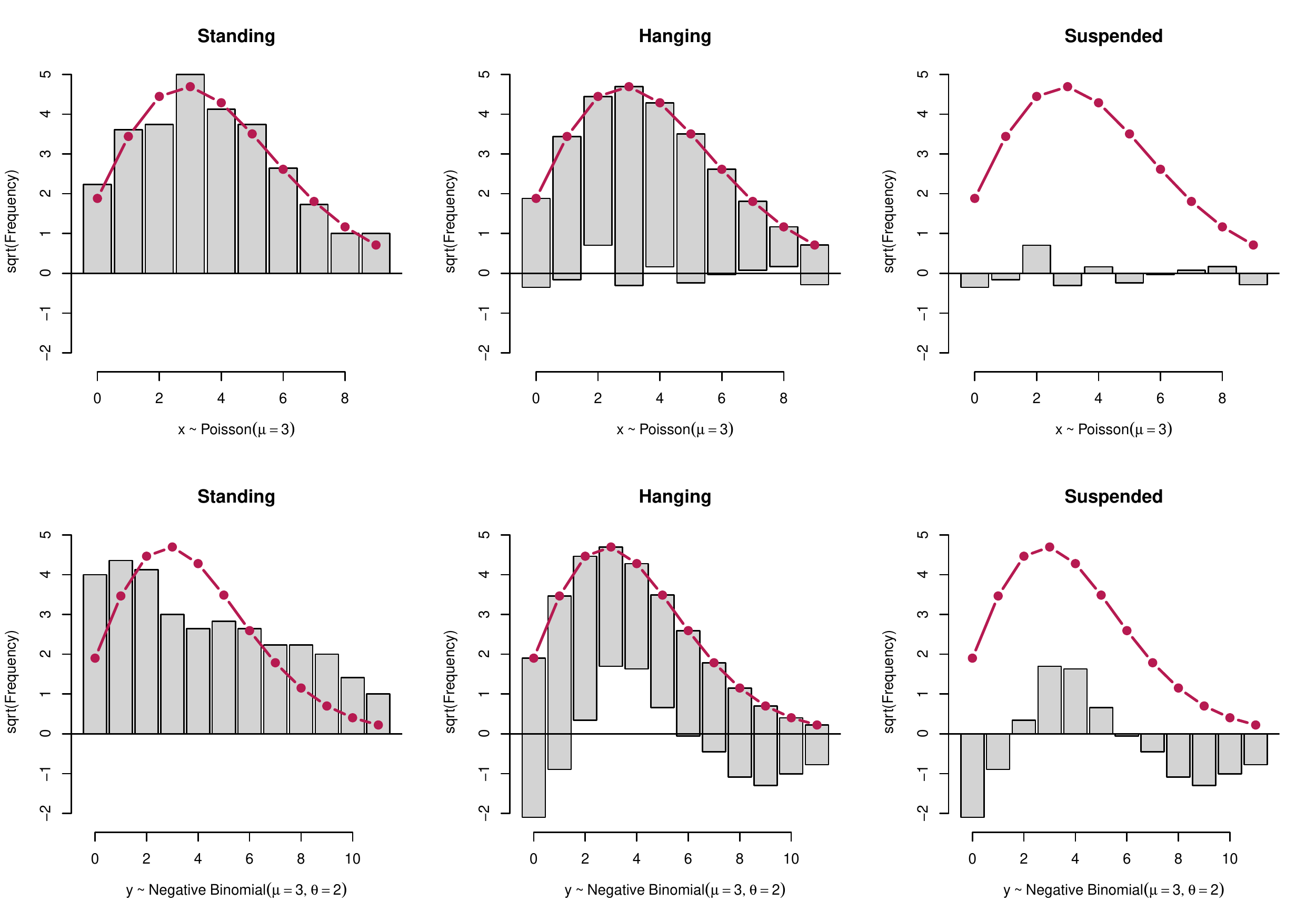}
\caption{\label{fig:styles} Styles of rootograms for two Poisson models fitted to 100 artificial
observations from a Poisson (top row) and negative binomial (bottom row) distribution. Top row:
The Poisson model fit ($\hat \mu = 3.34$) captures the true mean ($\mu = 3$)
as well as the distributional form with small deviations only. Bottom row: The Poisson model fit
($\hat \mu = 3.32$) does not capture the underlying distribution well
($\mu = 3$, $\theta = 2$), leading to clear deviations in the rootogram.}
\end{figure}

Different styles of rootograms have been suggested, see Figure~\ref{fig:styles}:
\begin{itemize}
  \item \emph{Standing:} The standing rootogram simply shows rectangles/bars for
    $\sqrt{\text{obs}_j}$ and a curve for $\sqrt{\text{exp}_j}$. To assess deviations
    across the $j$'s, the expected curve needs to be followed as the deviations
    are not aligned.
  \item \emph{Hanging:} To align all deviations along the horizontal axis, the rectangles/bars
    are drawn from $\sqrt{\text{exp}_j}$ to $\sqrt{\text{exp}_j} - \sqrt{\text{obs}_j}$
    so that they are ``hanging'' from the curve representing expected  frequencies, $\sqrt{\text{exp}_j}$.
  \item \emph{Suspended:} To emphasize mainly the deviations (rather than the observed
    frequencies), a third alternative is to draw rectangles/bars for
    the differences between expected and observed frequencies, $\sqrt{\text{exp}_j} - \sqrt{\text{obs}_j}$ 
    (some authors use $\sqrt{\text{obs}_j} - \sqrt{\text{exp}_j}$ instead).
\end{itemize}

The basic version, the standing rootogram, is perhaps the least useful among the
three: it simply plots rectangles/bars and a curve representing the model, but
the fit is not easily assessed. The other versions both make use of a horizontal
reference line, a detail often emphasized by Tukey \citep[e.g.,][]{rootograms:Tukey:1972}.  
Here, it highlights the discrepancies between observed and expected
frequencies. In a sense, hanging rootograms emphasize the fitted values and
suspended rootograms the corresponding residuals. We recommend the hanging
version as the default as long as residuals are not of main concern, and hence
employ hanging rootograms below. We also note that the suspended rootogram exists in several versions;  
in \cite{rootograms:Tukey:1972} it was turned upside down, i.e., with a curve resembling expected values 
below the bars resembling residuals. Here we follow \cite{rootograms:Friendly:2000}.

\subsection{Interpreting Rootograms} \label{subsec:interpreting}

In analyses employing rootograms, one is often interested in detecting patterns
such as runs of positive or negative deviations, which highlight aspects of the
model fit that might require further attention. For example, Figure~\ref{fig:styles} 
presents rootograms for a Poisson model fitted to two simulated data sets from a 
Poisson (top row) and negative binomial (bottom row) distribution. Both underlying
distributions have mean $\mu = 3$, the negative binomial has a shape parameter $\theta = 2$
(while the Poisson is formally a negative binomial with $\theta =\infty$). 

When fitting a Poisson model to the Poisson data, all three versions
of the rootogram in the top row exhibit only small deviations: 
The standing version shows that the curve representing expected frequencies closely tracks the histogram representing observed frequencies, 
there are also no clear patterns in the hanging and suspended versions. All this indicates that the model fits well. 

In contrast, when fitting a Poisson model to the negative binomial data in the bottom row 
there are substantial departures of the model from the data: in the standing version, the curve representing expected frequencies does not track the observed frequencies, there are also discernible patterns in both the hanging and suspended variants. Specifically, 
for the latter the rootogram bars form a `wave-like' pattern around the horizontal reference line: the data exhibit too many small counts, 
notably zeros, as well as too many large counts for a Poisson model to provide an adequate fit. In summary, the patterns encountered 
in the bottom row of Figure~\ref{fig:styles} reflect a substantial amount of overdispersion that is not captured by the
fitted Poisson distribution.

The patterns seen in the bottom panel of Figure~\ref{fig:styles} are theoretically supported by 
results presented by \cite{rootograms:Mullahy:1997}, 
who shows that Poisson mixtures exhibit a larger number of zeros (compared with a Poisson null model) as well as more mass in the upper tails 
and less mass in the center of the distribution. Mullahy's results rely on earlier work of \cite{rootograms:Shaked:1980}, 
who shows that mixing will generally spread out a distribution (from the exponential family) towards its tails. 
The negative binomial distribution is a gamma mixture of the Poisson distribution, hence these arguments are directly relevant in the case at hand.
 
While excess zeros are strictly implied by overdispersion \citep[compare][Prop.~1]{rootograms:Mullahy:1997}, 
there also exist situations in practice where the number of zeros is so large that merely correcting for overdispersion via, 
e.g., a negative binomial model does not solve the problem. These tend to exhibit a spike at zero in graphical displays 
and are often best treated by fitting a two-part model. We shall encounter an example below.

\section{An Example from Ethology} \label{sec:example}

In this section we present an empirical illustration revisiting a well-known data set
from ethology, for which excess zeros and, more generally, overdispersion require treatment. 
We select models using information criteria, notably the BIC, 
and use rootograms for highlighting deficiencies of fitted models.

\cite{rootograms:Brockmann:1996} investigates horseshoe crab mating. The crabs arrive on the beach
in pairs to spawn. Furthermore, unattached males also come to the beach, crowd
around the nesting couples and compete with attached males for fertilizations.
These so-called satellite males form large groups around some couples while
ignoring others. \cite{rootograms:Brockmann:1996} shows that the groupings are not driven
by environmental factors but by properties of the nesting female crabs.
Larger females that are in better condition attract more satellites.

\citet[Chapter~4.3]{rootograms:Agresti:2013} reanalyzes these data, modeling the number of satellites 
using count data regression techniques. The main explanatory variable is the female
crab's carapace width, but its
color and spine condition are also considered in some analyses -- with the ordered
factors for color and spine condition often treated as numeric variables.
In his analysis, \cite{rootograms:Agresti:2013} starts out from a Poisson model
with the standard log link and then goes on to consider both Poisson and negative
binomial models with both log and identity links. He finds that among these the
negative binomial model fits best but also notes that further refinements might be possible,
e.g., by allowing for zero inflation.

\begin{figure}[t!]
\centering
\includegraphics{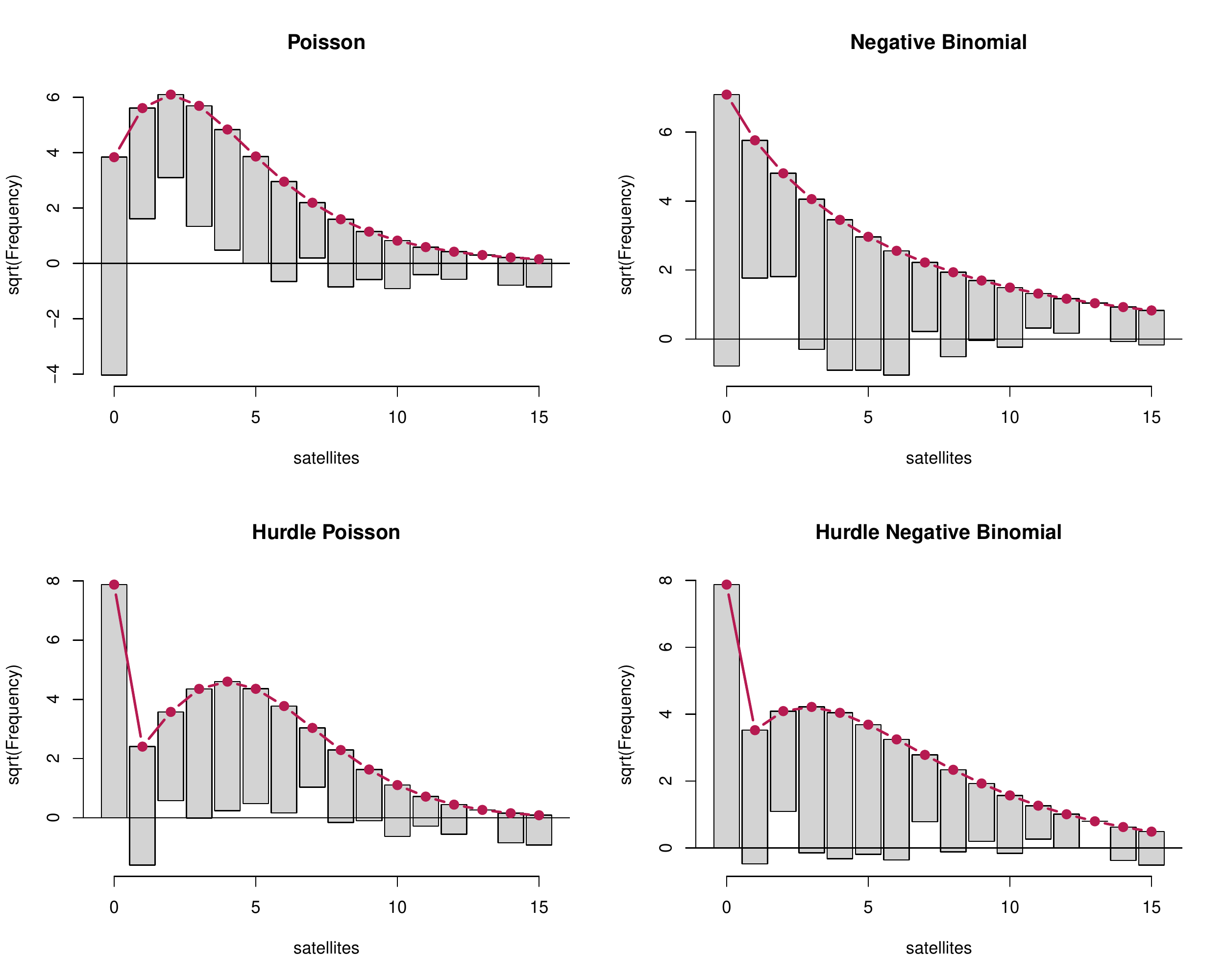}
\caption{\label{fig:CrabSatellites-rootograms} Hanging rootograms for crab
satellite models (counts $0, \dots, 15$).}
\end{figure}

To illustrate how rootograms can help in judging the goodness of fit of various count regression
models for this data, we extend the analysis of \cite{rootograms:Agresti:2013} in the following way:
we consider both Poisson and negative binomial regressions (with log link) and hurdle versions
of these (with a logit-type binary part) to allow for excess zeros. The carapace width and a numeric
coding of the color variable are used as regressors in all (sub-)models. To compare the relative
performances of the four models, we employ the Bayesian information criterion (BIC), yielding:
Poisson
(BIC = 931.0, df = 3),
negative binomial
(BIC = 769.5, df = 4),
hurdle Poisson
(BIC = 755.1, df = 6),
and hurdle negative binomial
(BIC = 736.8, df = 7).
These results already suggest that the hurdle negative binomial model fits best. However, a look at the 
corresponding
hanging rootograms (for counts $0, \dots, 15$) in Figure~\ref{fig:CrabSatellites-rootograms} provides much 
more 
insight into the pros and cons of the various models:
\begin{itemize}
  \item \emph{Poisson:} The wave-like pattern in the rootogram bars in the top left panel shows
    that the counts 1, \dots, 4 are overfitted while 0 and most counts from 6 onwards are
    underfitted. This indicates a substantial amount of overdispersion in the data, the clear lack of fit
    for~0 could be an additional indication of excess zeros.
  \item \emph{Negative binomial:} The rootogram does no longer exhibit the wave-like pattern of
    the Poisson model, showing that the overdispersion is accounted for much better in this
    model. However, the underfitting of the count~0 and clear overfitting for counts~1 and~2
    is typical for data with excess zeros. Note also that the fitted negative binomial model 
    implies a decreasing probability mass function (with $\theta = 0.93$), 
    which is not in line with the data structure.  
  \item \emph{Hurdle Poisson:} The rootogram now shows a perfect fit for the count~0 (by design
    of the hurdle model). However, there is still overdispersion in the remaining positive counts 
    that is again reflected by a wave-like pattern, note also the clear underfitting of the
    count~1.
  \item \emph{Hurdle negative binomial:} The rootogram shows that this model fits the data
    quite well. There are no clear patterns of departure anymore and the deviations between
    observed and predicted frequences are very small for most of the counts.
\end{itemize}

\begin{figure}[p!]
\centering
\includegraphics{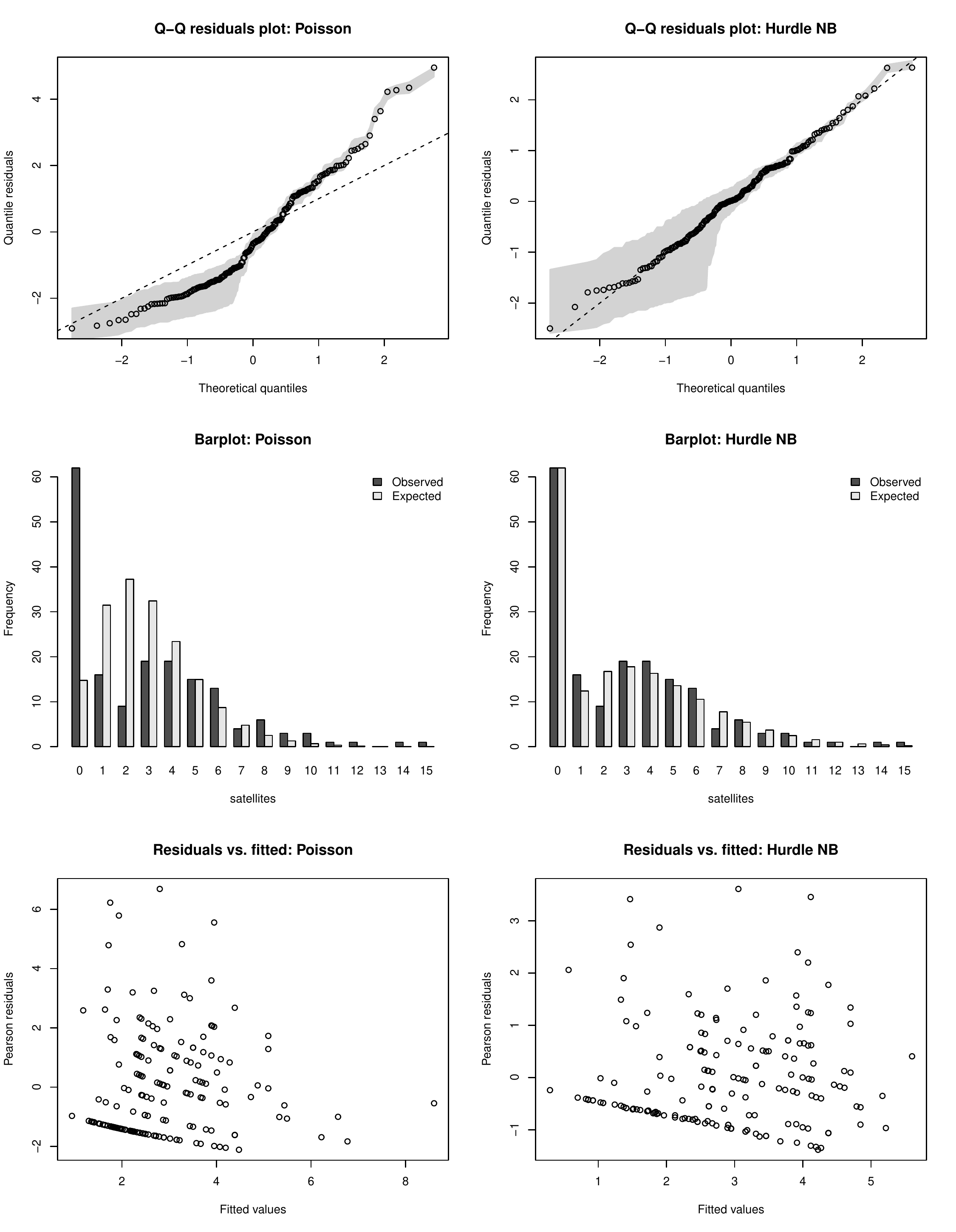}
\caption{\label{fig:CrabSatellites-comparison} Alternative graphical model checks for the crab satellites data.
Rows: Q-Q plot based on randomized quantile residuals, barplot of observed and fitted frequencies, and scatterplot of
Pearson residuals vs.\ fitted values (means). Columns: Poisson model and negative binomial hurdle model.}
\end{figure}

To highlight that the conclusions above are drawn more easily based on the proposed rootograms
than from more traditional visualizations, Figure~\ref{fig:CrabSatellites-comparison} provides different
types of displays for the poorly fitting Poisson model (left column) and the well-fitting
hurdle negative binomial (NB) model (right column). More traditional analyses include visualizations
of some sort of residuals, for example using quantile-quantile (or Q-Q) plots or plots of residuals vs. fitted values, 
and also of observed vs. expected frequencies. All three versions are used in sources such as \cite{rootograms:Cameron+Trivedi:2013}. 
The rows of Figure~\ref{fig:CrabSatellites-comparison} show:
\begin{enumerate}
  \item Quantile-quantile (or Q-Q) plots of randomized quantile residuals \citep{rootograms:Dunn+Smyth:1996}
    vs.\ the corresponding theoretical standard normal quantiles -- along with a gray shaded area corresponding to
    the range from the 5\% up to the 95\% quantile of the randomized distribution. The curvature of
    the Poisson model clearly indicates overdispersion but the excess zeros are not directly visible.
    The hurdle NB model, on the other hand, appears to fit rather well.
  \item Barplots of observed and expected frequencies. The excess zeros in the Poisson model are rather
    obvious while the overdispersion is somewhat obscured due to the tiny frequencies 
    of the larger counts. Also, the deviations are not aligned and hence are more difficult
    to track than in the hanging rootogram.
  \item Scatterplots of Pearson residuals vs.\ fitted values (means). Such displays are generally more
    difficult to interpret than in linear regression models due to the discrete and asymmetric response
    distribution. While it can be seen that the fit for the Poisson is not as good as for the hurdle NB
    model, the overall quality is harder to judge than in the previous displays. For the same reason,
    a scatter plot of observations vs.\ fitted means (not shown) would not be straightforward to
    interpret.
\end{enumerate}
Overall, rootograms clearly bring out several aspects that are not as easily seen in traditional displays. 
The barplots of observed and expected frequencies are closest in spirit to the rootogram, but suffer from an 
overemphasis of the tails (addressed by the square-root transformation in the rootogram) and the curved shape 
of the mass functions 
(addressed by the special alignment of observed vs. expected values and the horizontal reference line in the rootogram).

\begin{table}[t!]
\centering
\caption{\label{tab:CrabSatellites} Negative binomial hurdle models for crab
satellites. Coefficient estimates (and standard errors in parentheses).}

\medskip
%
%
\begin{tabular}{lD{.}{.}{3}D{.}{.}{3}cD{.}{.}{3}D{.}{.}{3}}
\toprule
&\multicolumn{2}{c}{Hurdle NB, model 1}&&\multicolumn{2}{c}{Hurdle NB, model 2}\\
\cmidrule{2-3}\cmidrule{5-6}
&\multicolumn{1}{c}{count}&\multicolumn{1}{c}{zero}&&\multicolumn{1}{c}{count}&\multicolumn{1}{c}{zero}\\
\midrule
(Intercept)&0.43&-10.07&&1.47&-10.07\\
&(0.94)&(2.81)&&(0.07)&(2.81)\\
width&0.04&0.46&&&0.46\\
&(0.03)&(0.10)&&&(0.10)\\
color&0.01&-0.51&&&-0.51\\
&(0.09)&(0.22)&&&(0.22)\\
Log(theta)&1.53&&&1.50&\\
&(0.35)&&&(0.35)&\\
\midrule
N&173&&&173&\\
Log-likelihood&-350.4&&&-351.0&\\
AIC&714.7&&&712.1&\\
BIC&736.8&&&727.8&\\
\bottomrule
\end{tabular}\end{table}

To further explore, the well-fitting hurdle NB model, its parameter estimates (and standard errors)
are reported in the first two columns of Table~\ref{tab:CrabSatellites}.
Interestingly, this reveals that the female crab's
carapace width and color both clearly affect the probability of having any satellites (binary
zero hurdle part). Specifically, larger crabs are much more likely to have satellites.
However, given that there is at least one satellite neither carapace width nor color are
individually significant (zero-truncated count part). Omitting both variables
improves the fit in terms of both AIC and BIC (hurdle NB, model 2, Table~\ref{tab:CrabSatellites}).
The rootogram of the simplified model (see Figure~\ref{fig:CrabSatellites-boot}) is very similar
to that of the full hurdle model.

\begin{figure}[t!]
\centering
\includegraphics{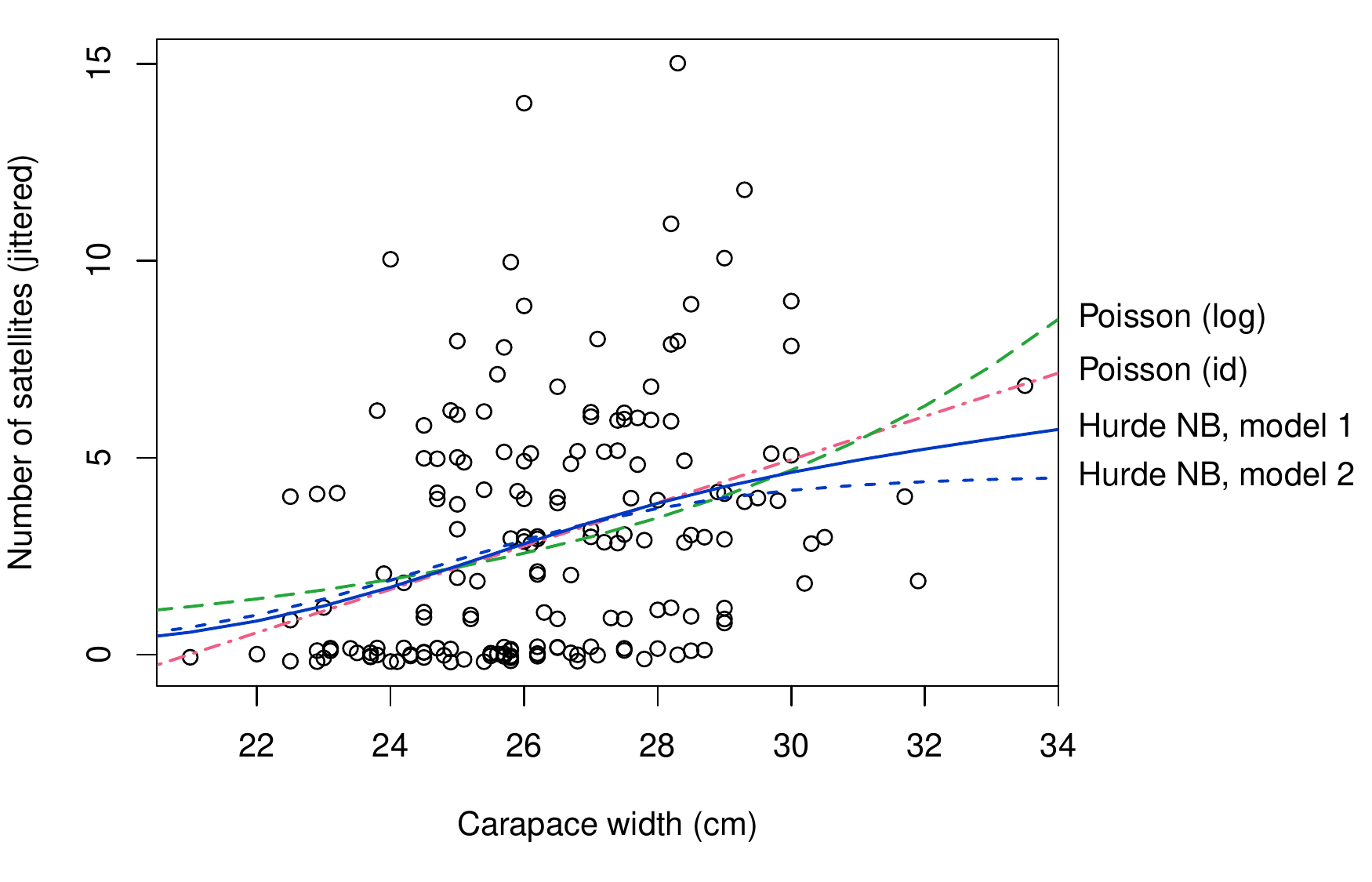}
\caption{\label{fig:CrabSatellites-effects} Predicted effect for the mean number of satellite
  at increasing carapace width and mean color.}
\end{figure}

Additionally, identity (rather than log) links or a zero-inflation (rather than hurdle) specification
could be employed but are omitted here for compactness. Both lead to qualitatively identical
insights and similar patterns in the rootograms while neither leads to improvements over the negative
binomial hurdle model. We conclude with a comparison of predicted effects for the mean function from several models.
(Figure~\ref{fig:CrabSatellites-effects}), evaluated for increasing carapace width at the mean color
(= 2.5 in the center of the scale 1, \dots, 4).
This shows that, compared to the identity link model preferred by \cite{rootograms:Agresti:2013},
the hurdle model leads to very similar predictions at average widths while avoiding negative predictions
for small widths and at the same time increasing even more slowly for large widths. This complements the
findings from the rootograms and underlines that the hurdle model fits the data rather well.

\setkeys{Gin}{width=0.7\textwidth}
\setkeys{Gin}{width=\textwidth}

\pagebreak

\section{Discussion and Concluding Remarks} \label{sec:disc}

Several flavors of rootograms have been discussed as graphical diagnostic tools for
visualizing complex regression models for count data. 
They combine exploratory data analysis with model-based
inference by bringing out discrepancies between observed and fitted distributions. 
Unlike other model-based graphics that often focus on effects on the mean of the fitted
distribution (e.g., effect displays), 
rootograms capture deviations across the support of the entire distribution and hence
can help to diagnose misfit regarding scatter and/or shape. This is particularly relevant
for count data models, which are often affected by problems such as
overdispersion and/or excess zeros.

To incorporate rootograms into the model-building workflow for count data regression models, their
graphical information can be used to complement standard techniques such as information criteria (AIC, BIC, \dots).
Using a range of basic models -- as done in Figure~\ref{fig:CrabSatellites-rootograms} for the crab
satellites data -- rootograms can guide the practitioner in deciding whether overdispersion
(e.g., Poisson vs.\ negative binomial models) and/or extra zeros (e.g., hurdle or zero-inflation
vs.\ `one-part' models) are relevant issues in the data at hand.
The models upon which the rootograms are based should use a reasonable first selection
of regressors (e.g., a standard specification from the literature or a model involving 
all potentially relevant variables). 

\begin{figure}[t!]
\centering
\includegraphics{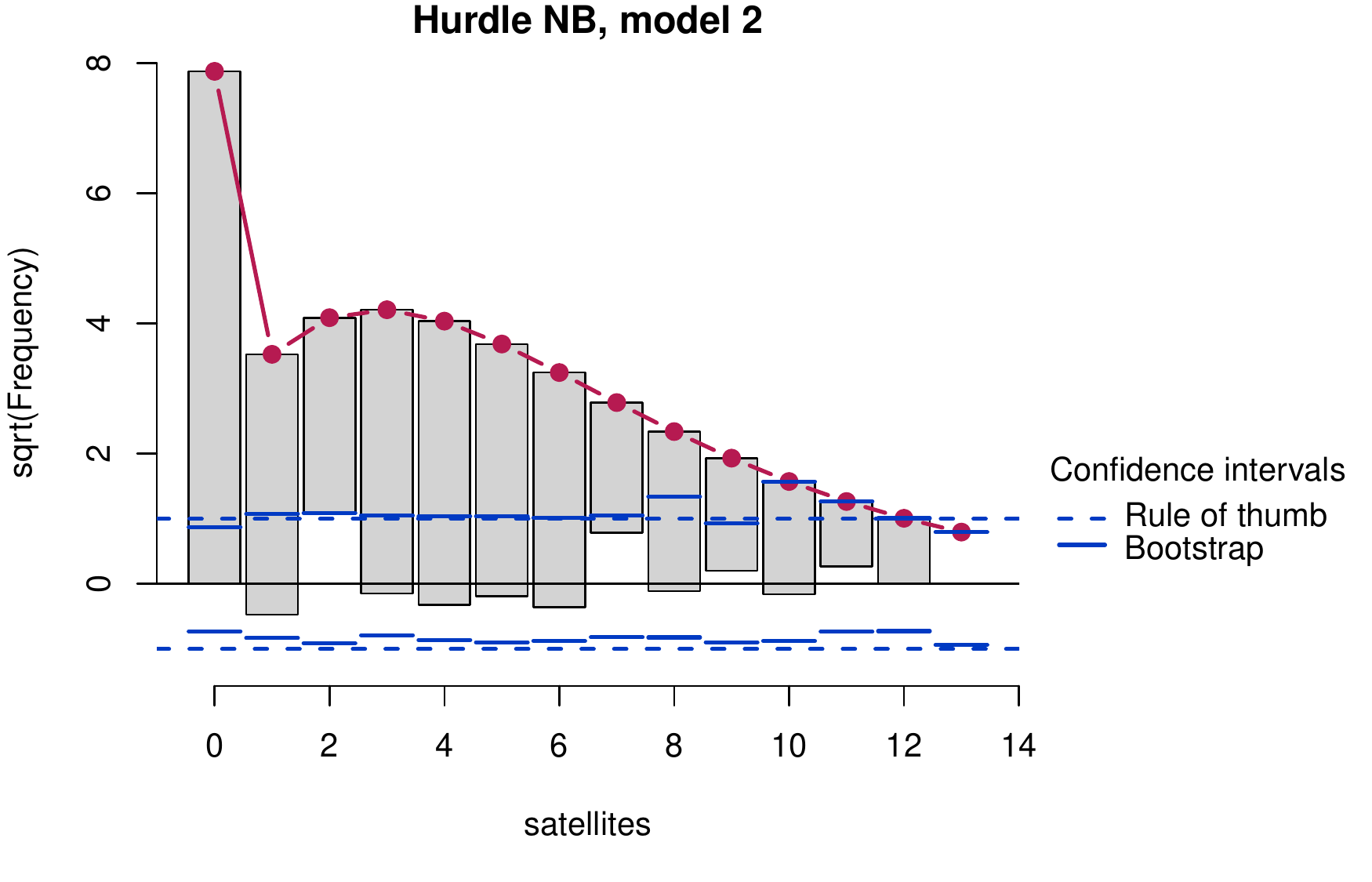}
\caption{\label{fig:CrabSatellites-boot} Hanging rootogram for the second hurdle NB model.
  Pointwise confidence intervals are added based on a rule of thumb ($\pm 1$, blue dashed)
  or the 2.5\% and 97.5\% quantiles from a parametric bootstrap (10,000 replications from the
  estimated model, blue solid).}
\end{figure}

Some users may want to include a built-in calibration of uncertainty. To this end, 
Figure~\ref{fig:CrabSatellites-boot} provides the rootogram of the simplified model 
from Table~\ref{tab:CrabSatellites} along with (pointwise) 95\%~confidence intervals obtained via 
a parametric bootstrap based on 10,000 replications from the estimated model. Note that the 
resulting intervals are not substantially different from the ``warning limits'' of \citet[][p.~61]{rootograms:Tukey:1972}, 
set at $\pm 1$, hence the latter would seem to be a useful practical device at a minimum cost. 

\setkeys{Gin}{width=0.7\textwidth}
\setkeys{Gin}{width=\textwidth}

Two further examples are available via supplements: the first uses a large data set from health economics, 
available from the \proglang{R} package \pkg{AER} supplementing \cite{rootograms:Kleiber+Zeileis:2008} under the name
\code{NMES1988} and exhibiting a 
different type of unobserved heterogeneity, the second involves the less frequent case of underdispersion and is available 
from the \proglang{R} package \pkg{countreg} \citep{rootograms:Zeileis+Kleiber:2016} under the name 
\code{TakeoverBids}. See the appendix below for more information on the latter package.

\section*{Computational Details} \label{sec:computational}

Our results were obtained using
\proglang{R}~3.2.4 \citep{rootograms:R:2016}
with the packages
\pkg{countreg}~0.1-5
\citep{rootograms:Zeileis+Kleiber:2016,rootograms:Zeileis+Kleiber+Jackman:2008},
\pkg{MASS}~7.3-45
\citep{rootograms:Venables+Ripley:2002}, and
\pkg{flexmix}~2.3-13
\citep{rootograms:Leisch:2004,rootograms:Gruen+Leisch:2008}, and 
were identical on various platforms including PCs running Debian GNU/Linux (with
a 3.2.0-1-amd64 kernel) and Mac OS X, version~10.10.5.

%
%
%

\section*{Acknowledgments} \label{sec:acknowledgments}

The authors thank the editor, the associate editor, and several anonymous reviewers for 
many valuable suggestions on earlier versions of this paper.

\bibliography{rootograms}

\newpage

\begin{appendix}

\section[R Implementation]{\proglang{R} Implementation}

For an overview of count data regression models in \proglang{R} we refer to 
\cite{rootograms:Zeileis+Kleiber+Jackman:2008}, where \proglang{R} implementations of hurdle 
and zero-inflation models are described in some detail. The corresponding fitting functions 
have now been moved to the \pkg{countreg} package, a new package that is 
currently under development by the authors of the present paper. First versions are already 
available from \url{http://R-Forge.R-project.org/projects/countreg/}.

The current implementation of rootograms in \pkg{countreg} provides
a generic function\linebreak \code{rootogram(object, ...)} along with several methods for
different types of models/data. The methods all proceed in the same way: They
first compute the observed and expected frequencies, $\text{obs}_j$ and $\text{exp}_j$
respectively (see Section~\ref{sec:rootograms}), and then call the default method
that computes all required coordinates for drawing the rootograms. The latter has
the following arguments:
\begin{Code}
  rootogram(object, fitted, breaks = NULL,
    style = c("hanging", "standing", "suspended"),
    scale = c("sqrt", "raw"), plot = TRUE,
    width = NULL, xlab = NULL, ylab = NULL, main = NULL, ...)
\end{Code}
The arguments \code{object} and \code{fitted} need to provide the tables/vectors
of observed and fitted frequencies. (The first argument is called \code{object}
rather than \code{observed} for consistency with the generic function that only
takes one required \code{object} argument and \code{...}.) The \code{breaks} need
to be specified if a continuous distribution is employed while for a discrete
distribution one may want to set the \code{width} of the bars to leave small gaps
between the bars (as in our examples). Additionally, one of three \code{style}s
can be specified: \code{"hanging"} (default), \code{"standing"}, or \code{"suspended"}.
The object returned is then a `\code{data.frame}' with all the coordinates needed for
plotting, and this is also drawn directly by default (\code{plot = TRUE}) along with
the specified graphical arguments (\code{xlab}, \code{ylab}, \code{main}, \code{...}).
By default, the base graphics \fct{plot} method is used for drawing rootograms. 
In addition, there is also an \fct{autoplot} method for drawing rootograms using the \pkg{ggplot2} package 
\citep{rootograms:Wickham:2009}.

Above we used methods for objects of classes `\code{glm}' and `\code{hurdle}'. There
are further methods available,  currently for univariate distributions fitted
via \fct{fitdistr} \citep[to objects of class `\code{numeric}',][]{rootograms:Venables+Ripley:2002},
zero-inflated models \citep[objects of class~`\code{zeroinfl}',][]{rootograms:Zeileis+Kleiber+Jackman:2008},
zero-truncated models (objects of class~`\code{zerotrunc}', as fitted by the \fct{zerotrunc} function in \pkg{countreg}), 
generalized additive models \citep[objects of class~`\code{gam}',][]{rootograms:Wood:2006},
and for selected count distributions falling within the framework of generalized
additive models for location, scale and shape
\citep[objects of class~`\code{gamlss}',][]{rootograms:Rigby+Stasinopoulos:2005,
rootograms:Stasinopoulos+Rigby:2007}. 




\section[Supplementary Material: Demand for Medical Care]{Supplementary Material: Demand for Medical Care} \label{sec:NMES1988}

Here we provide an additional example, taken from health economics. 
Its purpose is to present a larger data set with a much greater range of values for the response 
and further to show how fitted models resulting from modern tools such as finite mixture models can also be assessed via rootograms.

The data are cross-sectional data originating from the US National Medical
Expenditure Survey (NMES) conducted in 1987 and 1988. The NMES is
based upon a representative, national probability sample of the
civilian non-institutionalized population and of individuals admitted
to long-term care facilities during 1987. The subsample used here comprises only 
individuals aged 66 and over, all of whom are covered by
Medicare (a public insurance program providing substantial
protection against health-care costs). 
For \proglang{R} users, these data are conveniently available from the \pkg{AER}
package supplementing \cite{rootograms:Kleiber+Zeileis:2008} under the name
\code{NMES1988}. They have been explored
originally by \cite{rootograms:Deb+Trivedi:1997} using finite mixtures
of count data regressions. \cite{rootograms:Zeileis+Kleiber+Jackman:2008}
employ the data for illustration of hurdle and zero-inflation 
models while \cite{rootograms:Cameron+Trivedi:2013} reinvestigate
finite mixtures. Here, we follow the latter approach but employ a
slightly reduced set of regressors to facilitate interpretation
while still obtaining reasonably good fits.

\begin{figure}[t!]
\vspace*{-0.4cm}

\centering
\includegraphics{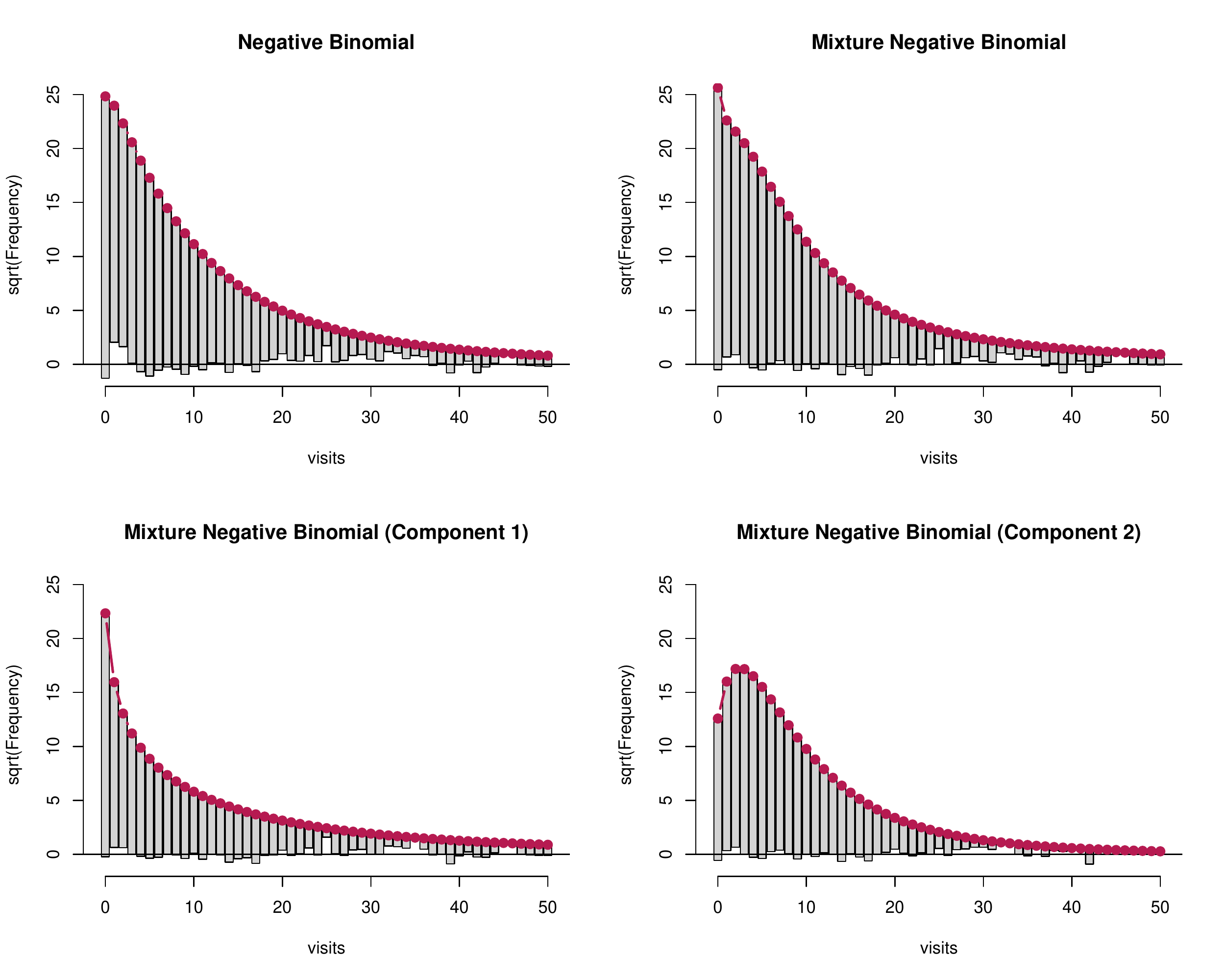}

\vspace*{-0.2cm}

\caption{\label{fig:NMES1988-rootograms} Hanging rootograms for NMES 1988 models.}
\end{figure}

Figure~\ref{fig:NMES1988-rootograms} displays the rootograms
for a single negative binomial regression as well as for a finite mixture
of two negative binomial regressions. For the latter, the mixture model (upper
panel, right) as well as both components (lower panel) are given.  The
corresponding parameter estimates (standard errors in parentheses) as well as the sums of the posterior weights (denoted $N$) are reported in Table~\ref{tab:NMES1988}.

\begin{table}[t!]
\centering
\caption{\label{tab:NMES1988} Negative binomial regression models (single and 2-component finite mixture)
  for NMES 1988 physician office visits. Coefficient estimates (and standard errors in parentheses).}

\medskip
%
%
\begin{tabular}{lD{.}{.}{3}cD{.}{.}{3}cD{.}{.}{3}}
\toprule
&\multicolumn{1}{c}{Single}&&\multicolumn{1}{c}{Component 1}&&\multicolumn{1}{c}{Component 2}\\
\midrule
(Intercept)&0.80&&-0.96&& 1.46\\
&(0.06)&&(0.39)&&(0.15)\\
health: poor/average&0.34&& 0.40&& 0.29\\
&(0.05)&&(0.14)&&(0.06)\\
health: excellent/average&-0.38&&-0.22&&-0.45\\
&(0.06)&&(0.16)&&(0.10)\\
chronic&0.19&& 0.27&& 0.16\\
&(0.01)&&(0.04)&&(0.02)\\
gender: male/female&-0.09&&-0.16&&-0.08\\
&(0.03)&&(0.09)&&(0.05)\\
school&0.03&& 0.07&& 0.01\\
&(0.00)&&(0.02)&&(0.01)\\
insurance: yes/no&0.35&& 1.70&&-0.10\\
&(0.04)&&(0.32)&&(0.10)\\
medicaid: yes/no&0.31&& 0.80&& 0.17\\
&(0.06)&&(0.26)&&(0.08)\\
\midrule
Log(theta)& 0.2&&-0.4&& 0.9\\
N&4406&&1744.9&&2661.1\\
Log-likelihood&-12215.0&&-12149.8&&\\
AIC&24448.0&& 24337.7&&\\
BIC&24505.5&& 24459.1&&\\
\bottomrule
\end{tabular}\end{table}

The single NB regression clearly misfits, especially for the low counts $0, 1,
2$, while the mixture NB provides an improved fit. It is possible to study the
mixture model in more detail by decomposing observed and expected frequencies
into the individual components and visualizing them separately. To this end, the observed
and expected frequencies are computed as weighted sums using the posterior
probabilities for each component. Figure~\ref{fig:NMES1988-rootograms} (bottom
panels) highlights nicely that both components fit rather well. It also brings
out the different means and variances in the two components. Specifically, the
first component contains a fraction of $0.4 =
1744.9/4406$ of all
observations and is characterized by a zero-modal rootogram. On average, the
corresponding individuals have fewer physician office visits but at the same
time a rather high variance. The parameter estimates are mostly larger (in absolute values) 
than in the second component, especially for the insurance and medicaid parameters. In
contrast, the second component is characterized by a unimodal rootogram with
comparatively lighter tails. On average, the corresponding individuals have more
physician office visits but at the same time a smaller variance. The first group
may, therefore, be seen as the group of occasional users, for which the number of visits
likely depends on the severity of the issues, while the second group may be seen
as the group of regular users, for which the number of visits often results from
the presence of chronic conditions. Indeed, when splitting the patients into two clusters 
(with hard assignment to the clusters according to the highest posterior probability),
it can be seen that the second cluster has
a lower proportion of persons with excellent health status
(10.1\% vs.\
 7.1\%),
or without chronic diseases
(34.4\% vs.\
 19.8\%),
and a higher proportion of insured persons
(64.7\% vs.\
 81.7\%).
Moreover, further unobserved factors such as the type of diseases and medication
might be captured by the two latent components.

\section[Supplementary Material: Takeover Bids]{Supplementary Material: Takeover Bids} \label{sec:TakeoverBids}

Our final example uses data from finance. Its purpose is to present
an application with underdispersion and fewer zeros than in the Poisson model.

The data comprise a set of firms that were targets of takeover bids during the period 1978--1985.
They were initially analyzed by \cite{rootograms:Jaggia+Thosar:1993} using a standard Poisson regression 
and are reanalyzed in \citet[][Chapter~5]{rootograms:Cameron+Trivedi:2013}. The response variable
is the number of takeover bids (after the initial bid received by the target firm) and a number
of regressor variables capturing the defensive actions of the target management and
firm-specific characteristics as well as potential interventions by federal regulators
are employed. A more detailed description of the variables is provided in 
\citet[][Table~5.2]{rootograms:Cameron+Trivedi:2013}. Coefficient estimates (and standard
errors) of the Poisson regression model are reported in the first column of Table~\ref{tab:TakeoverBids}.

\begin{table}[b!]
\centering
\caption{\label{tab:TakeoverBids} Poisson and hurdle Poisson for takeover bids data.
  Coefficient estimates (and standard errors in parentheses).}
\medskip
%
%
\begin{tabular}{lD{.}{.}{3}cD{.}{.}{3}cD{.}{.}{3}}
\toprule
&\multicolumn{1}{c}{Poisson}&&\multicolumn{1}{c}{Hurdle Poisson: count}&&\multicolumn{1}{c}{zero}\\
\midrule
(Intercept)&0.99&&1.14&&2.15\\
&(0.53)&&(0.76)&&(3.47)\\
legalrest: yes/no&0.26&&0.44&&0.97\\
&(0.15)&&(0.21)&&(0.98)\\
realrest: yes/no&-0.20&&-0.00&&-2.72\\
&(0.19)&&(0.25)&&(1.00)\\
finrest: yes/no&0.07&&0.27&&-1.47\\
&(0.22)&&(0.27)&&(1.17)\\
whiteknight: yes/no&0.48&&0.88&&1.19\\
&(0.16)&&(0.28)&&(0.87)\\
bidpremium&-0.68&&-1.35&&0.82\\
&(0.38)&&(0.53)&&(2.48)\\
insthold&-0.36&&-0.66&&-1.84\\
&(0.42)&&(0.61)&&(2.41)\\
regulation: yes/no&-0.03&&-0.06&&-1.14\\
&(0.16)&&(0.22)&&(0.98)\\
size&0.18&&0.24&&0.35\\
&(0.06)&&(0.07)&&(1.02)\\
size$^2$&-0.01&&-0.01&&0.01\\
&(0.00)&&(0.00)&&(0.18)\\
\midrule
N&126&&126&&\\
Log-likelihood&-184.9&&-159.5&&\\
AIC&389.9&&359.0&&\\
BIC&418.3&&415.7&&\\
\bottomrule
\end{tabular}\end{table}

\begin{figure}[t!]
\centering
\includegraphics{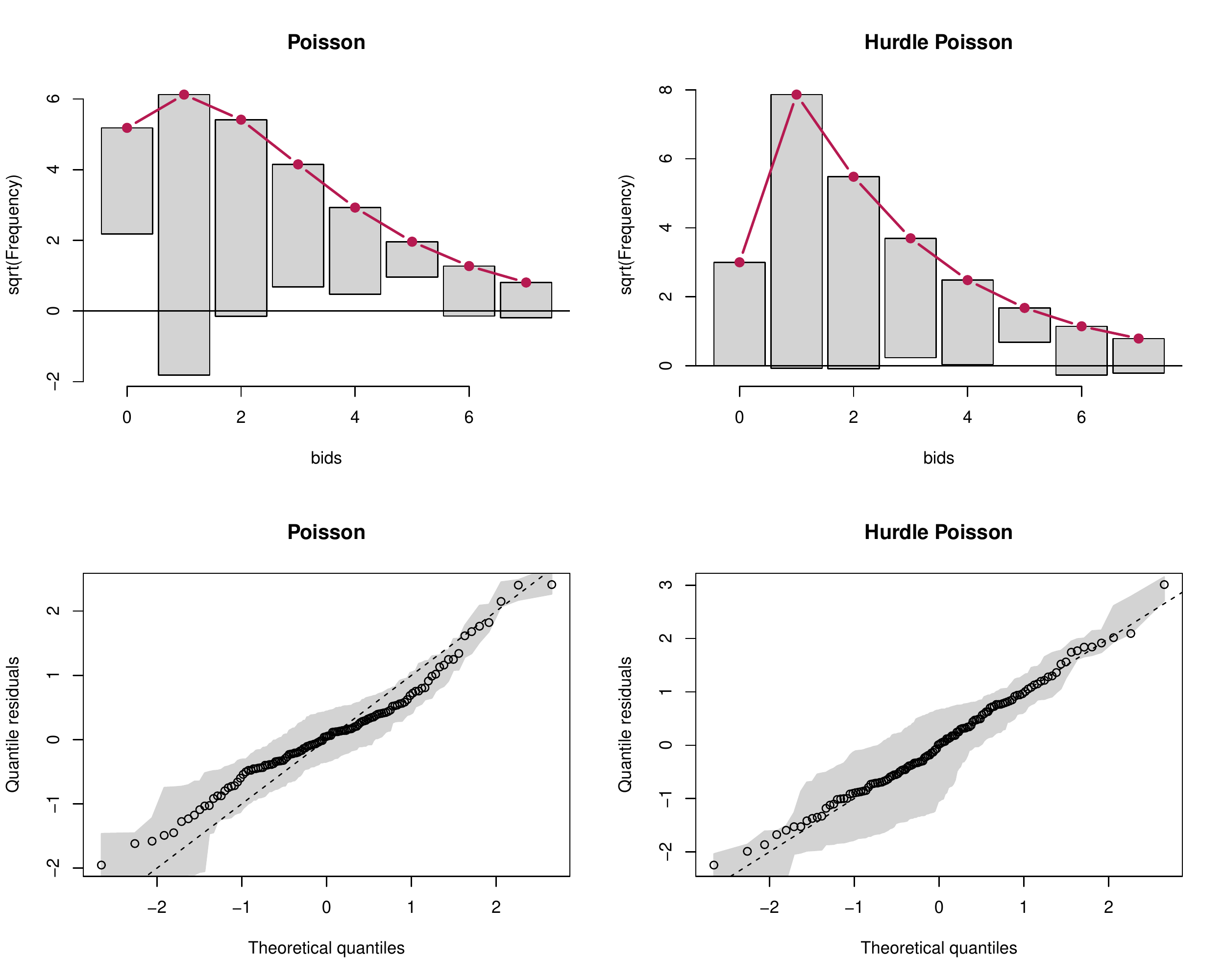}
\caption{\label{fig:TakeoverBids-plots} Hanging rootograms (top) and Q-Q residuals plots (bottom)
  for the Poisson (left) and hurdle Poisson (right) models for the takeover bids data.}
\end{figure}

The number of bids ranges from $0, \dots, 10$ so that a
Poisson model might capture the data well. However, only
7.1\% of the observations are zeros
which is fewer than expected under a Poisson model, leading to underdispersion in
the model. This is also brought out clearly by the corresponding hanging rootogram
in the top left panel of Figure~\ref{fig:TakeoverBids-plots}. One strategy to
improve the model is to employ a hurdle Poisson regression model, see the second
and third column of Table~\ref{tab:TakeoverBids}. This appropriately captures the
fewer zeros and leads to a satisfactory fit in the rootogram (top right panel of
Figure~\ref{fig:TakeoverBids-plots}).

In comparison, the corresponding Q-Q residuals plots (based on randomized quantile
residuals) also bring out the underdispersion in the Poisson model -- by way of the
curvature -- and  the satisfactory fit of the hurdle Poisson model. However, it
is less obvious that the underdispersion is mainly due to the fewer zeros in the data.

\end{appendix}

\end{document}